# The Abundance of Lead in Four Metal-Poor Stars

Ruth C. Peterson[1]

[1] SETI Institute, 189 Bernardo Ave, Suite 200, Mountain View, CA  94043

## Abstract

Cowan et al. (2021) review how roughly half the elements heavier than iron found in the Sun are produced by rapid neutron capture and half by slow neutron capture, the r- and s-processes. In the Sun, their relative contribution to individual elemental abundances is well understood, except for the lightest and heaviest elements beyond iron. Their contributions are especially uncertain for the heaviest non-radioactive element, lead (Pb, Z=82). This is constrained by deriving lead abundances in metal-poor stars. For in the most metal-poor halo stars, strontium and heavier elements are found in the solar r-process proportion; s-process elements appear only at metallicities above one-thirtieth solar.

In unevolved metal-poor stars of roughly solar heavy-element content, only two UV Pb lines are detectable. Four such stars have high-resolution spectra of the strongest line, Pb II at 2203.53Å. Roederer et al. (2020) analyzed this line in one star, deriving a lead-to-iron abundance ratio ten times solar. This and its blue-shifted profile suggested strong s-process production.

This work analyzes the UV spectra of all four stars. Calculations including a predicted Fe I line blueward of the Pb II line, and assuming the lead abundance scales with r-process abundances, match all four profiles extremely well. A scaled s-process contribution might improve the match to the much lower lead abundance found in the unevolved star analyzed previously, but its s-process excess is modest. An Fe II line blends the other lead line, Pb I at 2833.05Å, which constrains the lead abundance only in the coolest star.

**SECTION 1: Background.** The spectra of unevolved metal-poor halo stars uniquely reflect the elemental abundances incorporated during the earliest Galactic epoch. As reviewed by Sneden et al. (2008) and Cowan et al. (2021), their heavy-element content is understood generally as the products of the r- and s-process, rapid and slow neutron capture on iron-peak elements. While the ratio of r- to s-process products varies widely from star to star, and while some extremely metal-poor stars show extreme enhancement of the products of one or both processes, the distribution of abundances produced by each process remains relatively constant.

Until recently, supernovae explosions of the first massive stars formed were considered the most promising site of the r-process at early times. The merger of neutron stars/black holes must now be considered as well (Cowan et al. 2021); their evolution requires more time. The s-process also requires more time, for its main production site is in evolved asymptotic giant branch stars (Arlandini et al. 1999; Travaglio et al. 2004). Discerning the contributions from each process has largely been done via metal-poor stars; the oldest, most metal-poor such stars tend to have r-process but not s-process contributions (Sneden et al. 2008).

Lead (Pb; Z=82), the heaviest non-radioactive element, can be formed by both processes. Even in the Sun, lead has a poorly determined r-process contribution (Cowan et al. 2021, Fig. 2). Disentangling r- and s- contributions from stellar spectra can be challenging, because lead has few detectable lines. Optical Pb I lines are available for the Sun and cool, metal-poor giants, but all are very weak (e.g. Sneden et al. 1998; Plez et al. 2004).

Detection of lines of lead in spectra of stars with metallicity below one-tenth solar can be done in the UV, but lines remain weak and are affected by strong line blending. From the Pb I line at 2833Å, Sneden et al. (1998) derived an uncertain lead abundance for HD 115444 and Cowan et al. (2002) found an upper limit for BD +17 3248. Both stars are giants with extreme r-process enhancements, like many stars studied optically. Roederer et al. (2020) derived the lead abundance for a warm metal-poor turnoff star, HD 94028, from a Pb II line at 2203.5Å. Assuming that the feature is due to Pb II alone, they concluded that lead is enhanced by a factor of ten over its solar proportion with respect to iron. This and profile fitting led Roederer et al. (2020) to suggest that lead is heavily produced by the s-process, with significant ramifications for the site and timescale of lead production at early times.

Notably, two such warm turnoff stars that are even more metal-poor than HD 94028 have archival UV spectra that also show absorption at 2203.5Å. If due to Pb II alone, this feature also implies very large lead overabundances in these two stars, especially in the star HD 140283 that has lower-than-solar abundances of the r- and s-process elements relative to iron. Instead, the absorption at 2203.5Å in its spectrum could arise from an unidentified Fe I line, one whose upper energy is unknown. The uncertainty in its energy frequently causes Kurucz predicted wavelengths to be off by ≈10Å in the UV.

Many previously unidentified Fe I lines have been identified by Peterson & Kurucz (2015; PK15) and Peterson, Kurucz, and Ayres (2017; PKA17), by matching observed wavelengths of unidentified stellar spectral features to Kurucz quantum-mechanical predictions. However, numerous unidentified Fe I lines still occur in the UV, as Peterson et al. (2020; PBS20) showed and discussed in their analysis of five metal-poor turnoff stars.

**SECTION 2: Spectral Analysis.** This work analyzes the Pb II feature in all four of the high-resolution UV spectra available for metal-poor stars near the halo main-sequence turnoff. Strong UV blending restricts attention to the highest resolution UV spectra, those obtained with the Hubble Space Telescope STIS spectrograph and E230H grating. This analysis follows the procedure of PBS20, using the Kurucz program SYNTHE to calculate stellar spectra from first principles and then comparing them to archival E230H observations. PBS20 describes these procedures, model atmospheres, and input line lists.

Table 1 lists the four metal-poor turnoff stars with suitable existing STIS E230H spectra. After each star's HD number appear the parameters of the model atmosphere adopted for the calculation of its spectrum: effective temperature $T_{eff}$, gravity log g, [Fe/H] – the logarithm of the iron-to-hydrogen ratio with respect to that of the Sun, and microturbulent velocity $V_t$ in km s$^{-1}$. Next are [Ba/Fe] and [Eu/Fe], the corresponding logarithmic ratios of stellar barium and europium abundances to iron versus their solar ratios. [Ba/Fe] and [Eu/Fe] values provide estimates of s- and r-process enhancement respectively. For the top three stars, all entries are from PBS20: the stellar parameters are from the SYNTHE entries of Table 2, and [Ba/Fe] and [Eu/Fe] are from Table 5. HD 211998 entries are values updated from Peterson (2011), from a SYNTHE analysis that incorporated its E230H spectrum and an improved line list. The E230H spectra were obtained under program GO-14161 (PI Peterson), with grating settings c2263 and c2762. Other E230H archival E230H datasets have lower S/N (e.g. those of GO-7348 near 2200Å: PBS20).

**Table 1: Properties of Turnoff Stars**

| Star: HD | $T_{eff}$ | log $g$ | $V_t$ | [Fe/H] | [Ba/Fe] | [Eu/Fe] | [Pb/Fe] |
|---|---|---|---|---|---|---|---|
| 84937 | 6300 | 4.0 | 1.3 | –2.25 | –0.22 | +0.38 | +0.4 ± 0.2 |
| 94028 | 6050 | 4.3 | 1.3 | –1.40 | +0.17 | +0.15 | +0.3 ± 0.2 |
| 140283 | 5700 | 3.6 | 1.3 | –2.60 | –0.77 | –0.22 | <+0.0 |
| 211998 | 5300 | 3.3 | 1.6 | –1.60 | –0.10 | +0.20 | +0.2 ± 0.2 |

Note: $V_t$ units are km s$^{-1}$.

Table 1 also includes these UV results for lead. Their derivation is illustrated in Figures 1 – 3. Calculated spectra (thin lines) are superimposed on observed E230H spectra (heavy black lines) in the two UV wavelength regions containing lines of lead. Blue lines adopt [Pb/Fe] = [Eu/Fe], and purple lines show the effect of changing [Pb/Fe] alone by ± 0.3 dex. Each stellar comparison is offset vertically. Y-axis ticks are spaced 10% apart in residual intensity. X-axis ticks are spaced by 0.1Å in wavelength (in air). Above each comparison appears the stellar HD number, followed in Figs. 1 and 2 by the parameters of the model atmosphere adopted. At the top are identifications and parameters for strong calculated lines. After the decimal digits of wavelength are the species, lower excitation potential in eV, the decimal digits of the intrinsic core residual intensity, and log gf value.

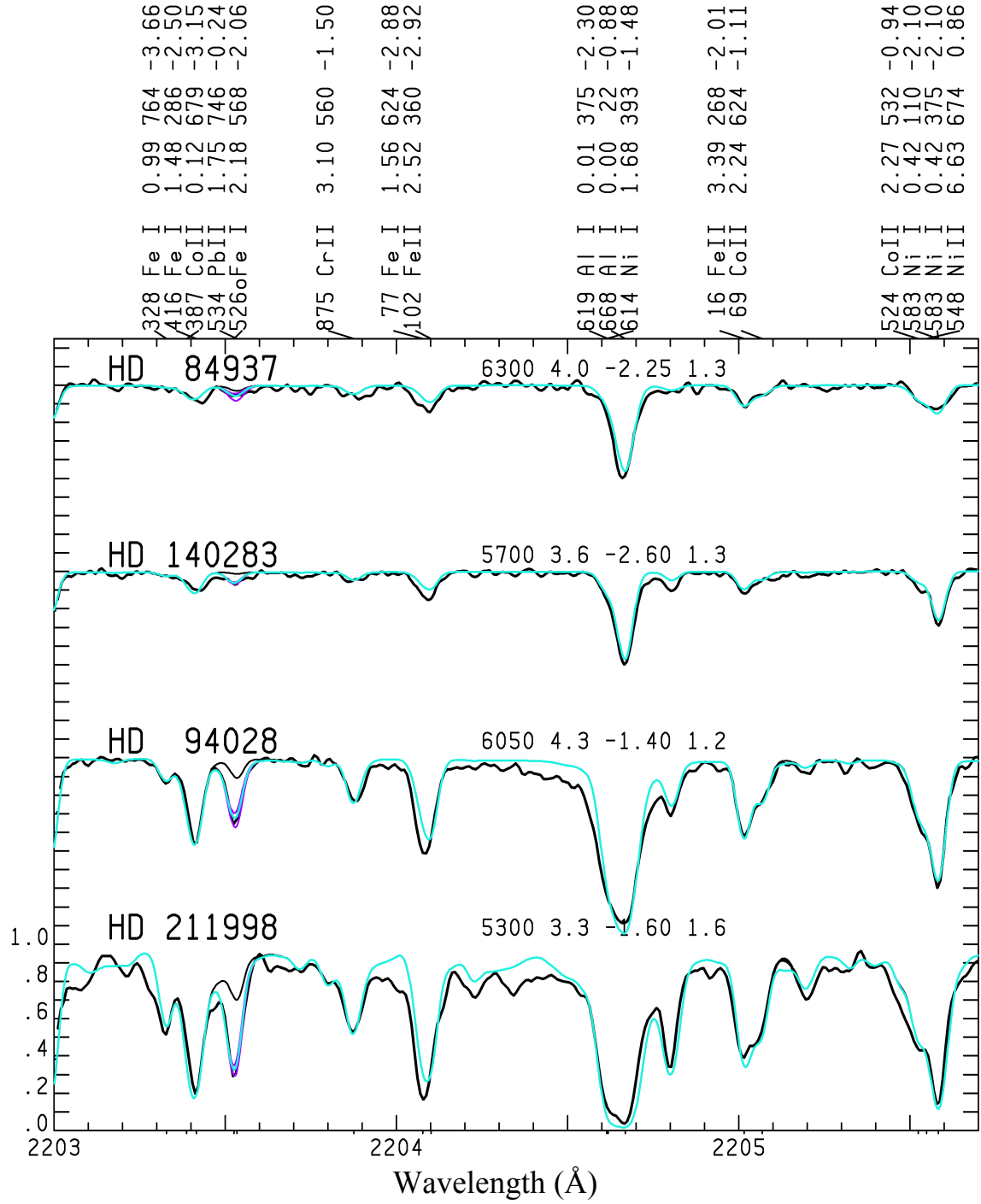

Figure 1. Four plots compare observed and calculated spectra near Pb II at 2203.53Å.

FIGURE 2

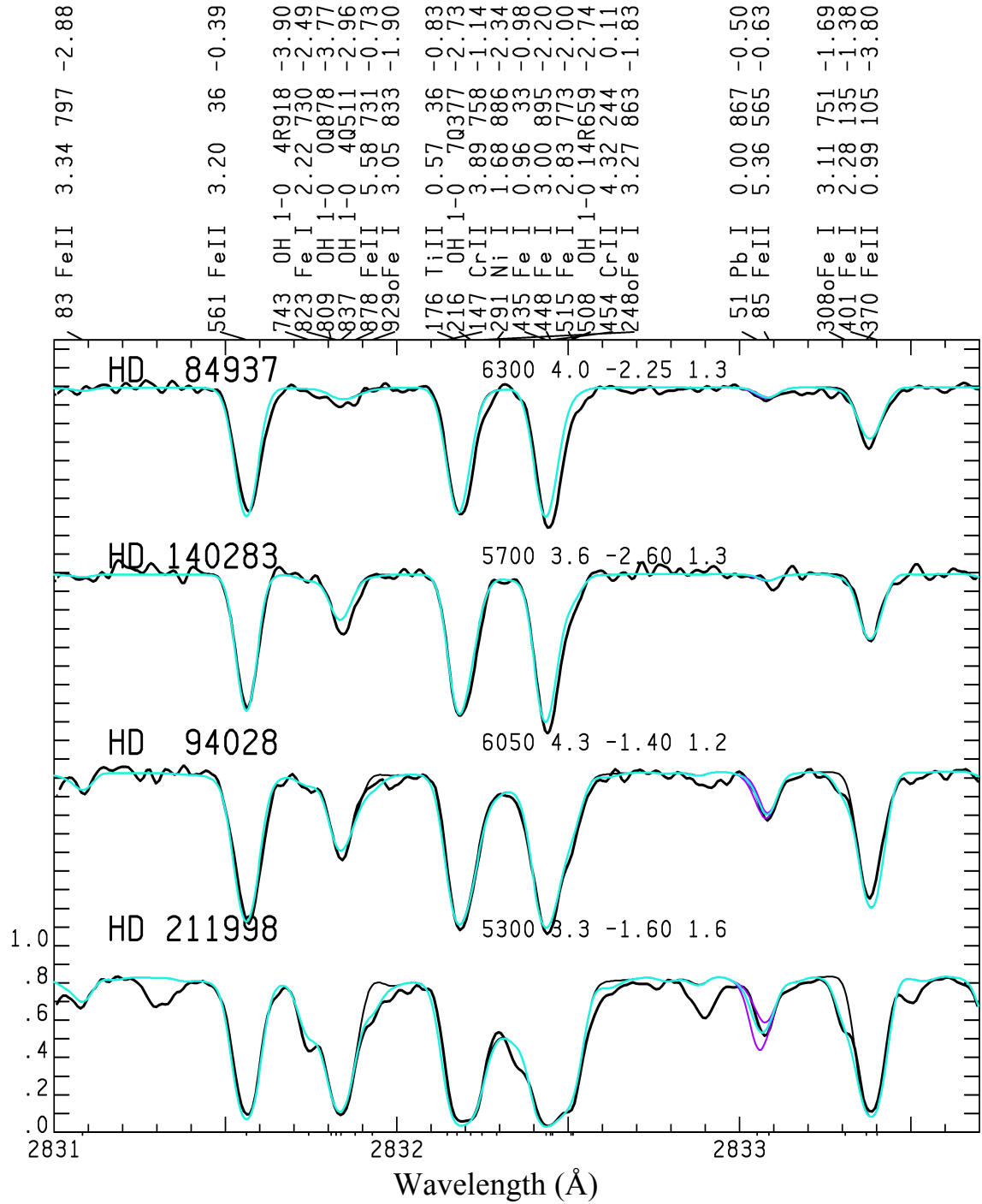

Figure 2. Plots similar to those of Fig. 1 compare spectra near Pb I at 2833.05Å.

FIGURE 3

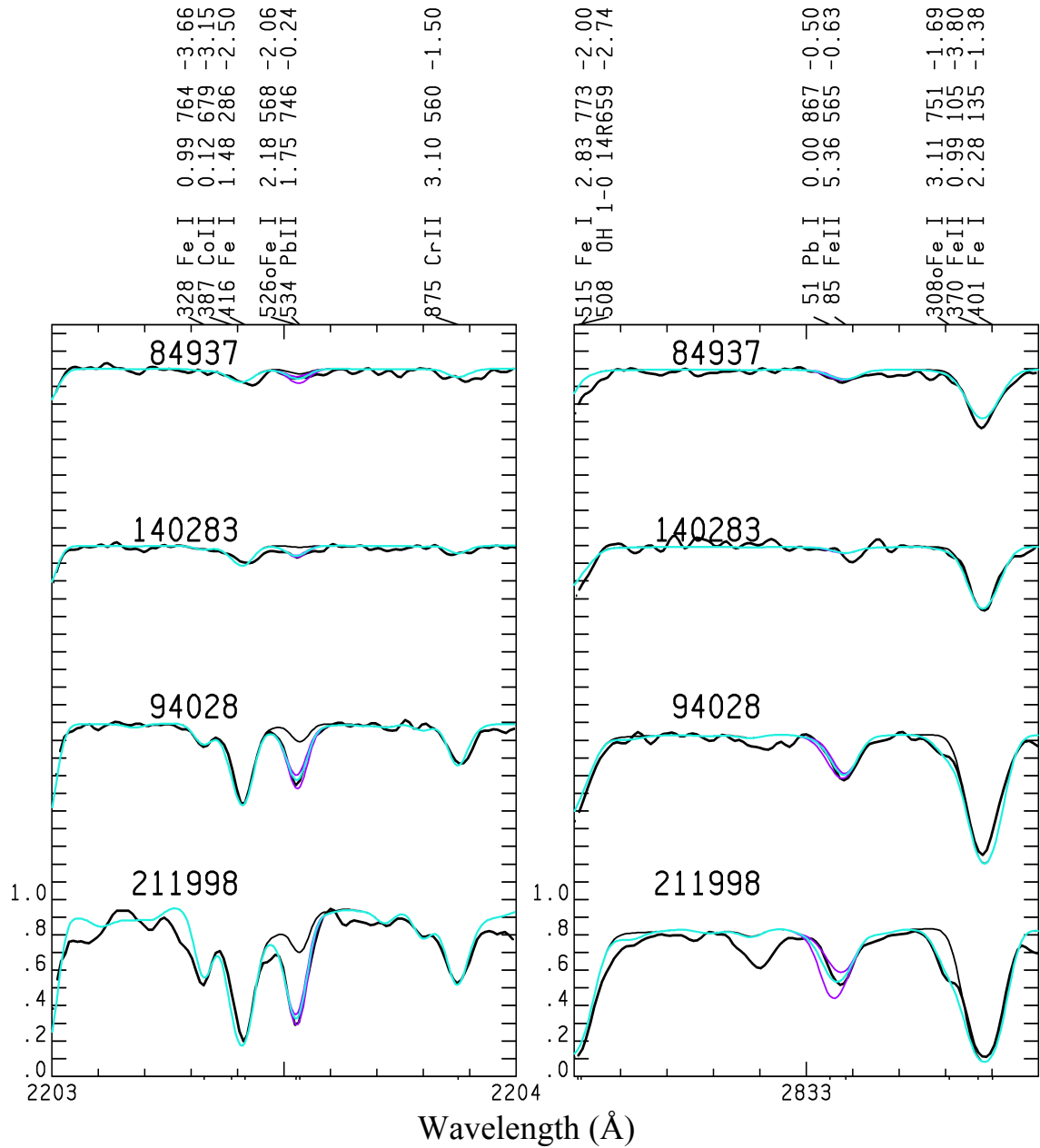

Figure 3. Expanded plots show regions of Figs. 1 and 2 containing Pb II and Pb I lines.

The lead abundance and its uncertainty given in Table 1 is estimated from the fits among the blue and purple profiles, as the latter show the effect of changing [Pb/Fe] by ±0.3 dex.

Unlike calculations for the black line, calculations for the light blue line also incorporate the newly identified Fe I lines of PK15 and PKA17, and Fe I lines marked o, each of which is an unconfirmed unidentified line whose true position is postulated to be at the wavelength whose decimal digits immediately precede its o. The effect of the newly identified and postulated lines on the spectral match is revealed wherever the light blue line falls below the thin black line.

Remaining unidentified lines are quite weak. They are best seen in the cooler HD 211998 spectrum, where the solid black line of the observed spectrum falls below the blue line.

In Figure 1, the thin black line near 2203.53Å in each comparison indicates the strength of the Pb II line itself; the blue line shows the additional effect of the postulated line. This line was drawn from the Kurucz file http://kurucz.harvard.edu/atoms/2600/gf2600.lines. In this file, a negative energy signifies an unknown energy and thus an uncertain wavelength. The file contains dozens of potentially detectable unidentified lines within 10Å of the Pb II line. A line with calculated wavelength = 2196.30Å and log gf = –1.66 was selected, due to its low level of excitation. A wavelength 2203.526Å and log gf = –2.06 were adopted. This identification is unconfirmed, for the implied wavelengths of the other two detectable lines of this level fall near the cores of much stronger features.

This postulated 2203.526Å line produces measurable absorption even in the top two spectra, most notably in HD 140283 where Pb II makes no contribution to the profile. Matching that profile by assuming it is due only to Pb II yields a lead abundance [Pb/Fe] = +0.7, +0.92 and +1.47 dex higher than its r- and s-process abundance levels (Table 1). This seems highly unlikely given that its r- and s-process abundance levels are subsolar, and that the Pb I 2833.051Å feature would be 4% deep but escaped detection in Fig. 2.

In HD 84937, both the Fe I and Pb II lines contribute to the 2203.53Å feature. Its profile is well matched assuming [Pb/Fe] = [Eu/Fe]. An excess ≥ +0.2 dex is ruled out.

The postulated 2203.526Å Fe I line dominates the Pb II feature in the two stronger-lined spectra. In the cool subgiant HD 211998, the line is sufficiently dominant and saturated that the lead abundance remains undefined. This dominance is also likely in cool giants.

For HD 94028, the fit to the 2203.53Å profile is marginally improved with the scaled r-process lead abundance increased by +0.17 dex, its s-process overabundance (Table 1). Adopting an r+s enhancement for lead yields [Pb/Fe] = +0.32, which is 0.63 dex lower than that found by Roederer et al. (2020) upon assuming the feature is due to Pb II alone.

The Pb I 2833.051Å feature is often considered an independent indicator of the lead abundance, both in giants and in dwarfs. However, Figure 2 shows that it is seriously blended by an Fe II line whose wavelength (in air) is well established at 2833.0852Å (Nave & Johansson 2013). Incorporating it with the log gf-value indicated in Fig. 2 again results in the blue line matching the 2833.05Å line profile in the three stars with E230H spectra of adequate S/N. In HD 211998, the lead abundance is constrained to that of the scaled r-process ±0.2 dex. In HD 84937 and HD 94028, Fe II dominates the Pb I profile, which barely constrains the lead abundance. Roederer et al. (2016) have provided a lead abundance determination for HD 94028 from this line, but their result is surely affected by the Fe II blend, especially since they adopted lower-resolution E230M spectra.

Lead abundances in 14 field metal-poor giants with [Fe/H] < −1.3, among which 13 have [Eu/Fe] < +0.5, were derived by Aoki & Honda (2008) and compared to lead abundances derived by Yong et al. (2006, 2008) for giants in globular clusters. Both studies relied on the optical Pb I line at 4058Å. Like this work, Aoki & Honda (2008) conclude that the lead production in these stars arises from the r-process. However, their Fig. 2 illustrates that these two studies both find [Pb/Eu] ~ −0.6, rather than this result that [Pb/Eu] ~ 0. This discrepancy may point to genuine differences in the proportion of the heaviest elements produced by the r-process.

Analytical problems cannot yet be ruled out, since different lines were used: gf-values and ionization state may be responsible, as well as extreme crowding at 4058Å (Yong et al. 2006, 2008; Plez et al. 2004; Sneden et al. 2008), which potentially affects continuum placement. Additional UV analyses of warm metal-poor stars spanning a range of mild r- and s-process enhancements might better constrain how lead production occurred at early times. This in turn could illuminate whether supernovae and neutron-star mergers are both required to explain early r-process production, and to what extent the mass range of the stellar protagonists or the time delays in their evolution played an influential role.

**SECTION 3: Summary.** This analysis yields abundances in three metal-poor stars that are consistent with the production of lead at early Galactic epochs by the r-process alone, perhaps assisted by the s-process in the one star that shows mild s-process excess. Further work is needed over a wider range of r- and s-process enhancements to discern their individual roles. Pb II studies are well advised to anticipate Pb II blending in spectra of cool metal-poor giants: the postulated Fe I line that blends Pb II has a low excitation potential that strengthens this Fe I line at cool temperatures. A Pb I line at 2833.051Å is useful in cool metal-poor turnoff stars, but in warmer ones is too blended by an Fe II line.

**Acknowledgements.** Monique Spite and Beatriz Barbuy are thanked for helpful discussions. The UV spectra incorporated in this analysis were obtained with the NASA/ESA Hubble Space Telescope under GO 14161 (PI Ruth Peterson). Additional support for this work under HST programs GO 14161 and GO 15179 (PI Ruth Peterson) was provided through NASA grants from the Space Telescope Science Institute, which is operated by the Association of Universities for Research in Astronomy, Inc., under NASA contract NAS 5-26555.

**REFERENCES**

Aoki, W., & Honda, S.  2018, PASJ, 60, L7

Arlandini, C., Käppeler, F., Wisshak, K., et al.  1999, ApJ, 525, 886

Cowan, J.J., Sneden, C., Lawler, J.E., et al. 2021, arXiv:1901.01410v3 [astro-ph.HE] 1 Feb 2021

Cowan, J.J., Sneden, C., Burles, S., et al.  2002, ApJ, 572, 861

Nave, G., & Johansson, S. 2013, ApJS, 204, 1

Peterson, R.C. 2011, ApJ, 742, 21

Peterson, R.C., & Kurucz, R.L. 2015, ApJS, 216, 1 (PK15)

Peterson, R.C., Kurucz, R.L., & Ayres, T.R. 2017, ApJS, 229, 23 (PKA17)

Peterson, R.C., Barbuy, B., & Spite, M. 2020, A&A, 638, A64 (PBS20)

Plez, B., Hill, V., Cayrel, R., et al. 2004, A&A, 428, L9

Roederer, I.U., Karakas, S.I., Pignatari, M., & Herwig, F. 2016, ApJ, 821, 37

Roederer, I.U., Lawler, J.E., Holmbeck, E.M., et al. 2020, ApJL, 902, L24

Sneden, C., Cowan, J.J., Burris, D.L., & Truran, J.W. 1998, ApJ, 496, 235

Sneden, C., Cowan, J.J., & Gallino, R. 2008, ARA&A, 46, 241

Travaglio, C., Gallino, R., Arnone, E., et al. 2004, ApJ, 601, 864

Yong, D., Aoki, W., Lambert, D.L., & Paulson, D.B. 2006, ApJ, 639, 918

Yong, D., Lambert, D.L., Paulson, D.B., & Carney, B.W. 2008, ApJ, 673, 854